# Feature-enhanced Adversarial Semi-supervised Semantic Segmentation Network for Pulmonary Embolism Annotation

Ting-Wei Cheng[a], Jerry Chang[a], Ching-Chun Huang[b], Chin Kuo[c,d,‡], Yun-Chien Cheng[a,‡]

[a] *Department of Mechanical Engineering, College of Engineering, National Yang Ming Chiao Tung University, Hsin-Chu, Taiwan*

[b] *Department of Computer Science, College of Computer Science, National Yang Ming Chiao Tung University, Hsin-Chu, Taiwan*

[c] *Department of Oncology, National Cheng Kung University Hospital, College of Medicine, National Cheng Kung University, Tainan, Taiwan*

[d] *College of Artificial Intelligence, National Yang Ming Chiao Tung University, Hsin-Chu, Taiwan*

[‡]*The authors contributed equally to this work.*

*Corresponding author: yccheng@nycu.edu.tw , tiffa663@gmail.com

## Abstract

This study established a feature-enhanced adversarial semi-supervised semantic segmentation model to automatically annotate pulmonary embolism (PE) lesion areas in computed tomography pulmonary angiogram (CTPA) images. In current studies, all of the PE CTPA image segmentation methods are trained by supervised learning. However, the supervised learning models need to be retrained and the images need to be relabeled when the CTPA images come from different hospitals. Therefore, this study proposed a semi-supervised learning method to make the model applicable to different datasets by adding a small amount of unlabeled images. By training the model with both labeled and unlabeled images, the accuracy of unlabeled images can be improved and the labeling cost can be reduced. Our semi-supervised segmentation model includes a segmentation network and a discriminator network. We added feature information generated from the encoder of segmentation network to the discriminator so that it can learn the similarity between predicted mask and ground truth mask. The HRNet-based architecture was modified and used as the segmentation network. This HRNet-based architecture can maintain a higher resolution for convolutional operations so the prediction of small PE lesion areas can be improved. We used the labeled open-source dataset and the unlabeled National Cheng Kung University Hospital (NCKUH) (IRB number: B-ER-108-380) dataset to train the semi-supervised learning model, and the resulting mean intersection over union (mIOU), dice score, and sensitivity achieved 0.3510, 0.4854, and 0.4253, respectively on the NCKUH dataset. Then, we fine-tuned and tested the model with a small amount of unlabeled PE CTPA images from China Medical University Hospital (CMUH) (IRB number: CMUH110-REC3-173) dataset. Comparing the results of our semi-supervised model with the supervised model, the mIOU, dice score, and sensitivity improved from 0.2344, 0.3325, and 0.3151 to 0.3721, 0.5113, and 0.4967, respectively. In conclusion, our semi-supervised model can improve accuracy on other datasets and reduce the labor cost of labeling with only a small amount of unlabeled images for fine-tuning,.

## 1. Introduction

Pulmonary embolism (PE) is a disease in which the blood vessels in the lungs are blocked by foreign bodies, resulting in hypoxia in the lung tissue. Acute PE can even lead patient to death. Since the symptoms are similar to those of a cold, doctors need to arrange a variety of diagnosis to confirm the PE condition. The diagnosis include electrocardiogram, chest X-ray, computed tomography (CT) scan, blood test, and computed tomography pulmonary angiogram (CTPA). Among these diagnoses, CTPA is the main criterion for the final diagnosis of PE. Therefore, the detection of PE in CTPA images is clinically important. However, in PE CTPA detection, doctors need to find the lesion area from a large amount of images, which makes diagnosis difficult. In order to speed up the diagnosis and improve the PE CTPA detection accuracy, the use of deep learning to classify and annotate PE has become a popular research topic in recent years.

The deep learning research on PE CTPA images can be divided into two categories. First category is CTPA images classification, which is to find out patients with PE early to reduce the mortality. There are many studies on PE CTPA image classification [1-3] and the use of deep learning can achieve rapid classification. Another category is the PE lesion segmentation, which enables doctors to find the lesion area of PE quickly and easily and give follow-up treatment. Most of the current studies on the PE lesion segmentation with supervised learning [4, 5] encounter three challenges: (1) A great amount of manual labeling is required, resulting in huge labor costs. (2) The same datasets are used during both training and testing and overfitting occurred when using one single dataset. (3) Different hospitals have different machine settings where the PE CTPA imaging is also different. As a result, the accuracy may decrease when different datasets are used for testing. Therefore, as shown in Fig 1, our research aimed to build a model which can adapt to different datasets through semi-supervised learning methods. Semi-supervised learning uses both labeled images and unlabeled images for training, so multiple datasets can be used to fine-tune the model. In addition to reduce the need for a large amount of manual labeling, the semi-supervised model can also increase the model adaptability of different datasets through adding unlabeled images.

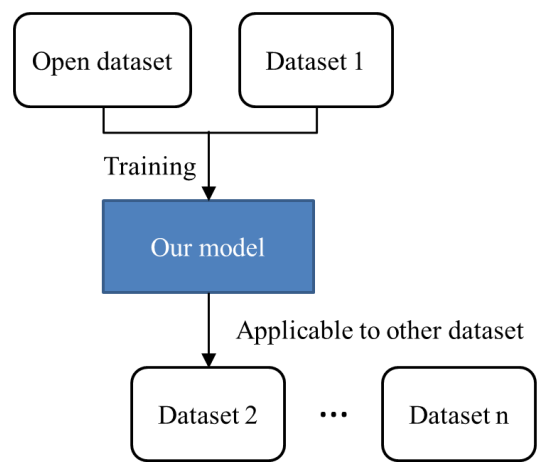

Fig 1. Application of our model. Our semi-supervised model was trained by an open-source dataset and an unlabeled dataset. The model can be fine-tuned by other unlabeled datasets to apply to different datasets.

We established a feature-enhanced semi-supervised segmentation model for annotating PE CTPA images. We trained the model through a labeled dataset and an unlabeled dataset, so that the model can adapt better on the unlabeled dataset. Then we used an external unlabeled dataset to fine-tune and test our model. Moreover, we strengthened the ability of our model in PE feature extraction. The improvement and advantages of our proposed model included:

- In the design of our model, we strengthened the ability of the discriminator by adding feature maps to the input of the discriminator during training, so that the predicted boundary and the shape of unlabeled PE lesion can be revised.

- HRNet-based architecture was used to enhance PE feature extraction. We reduced a downsampling layer and removed the 4[th] stage of HRNet [6] to reduce the calculation and extract PE features at a higher resolution. In additional, the bilinear upsampling layer in the decoder was replaced with the CARAFE [7] upsampling module to better restore the PE shape and position.

- This is the first study to establish semi-supervised segmentation for PE CTPA datasets and the training method can be applied to other PE CTPA datasets.

The use of semi-supervised learning method can reduce the demand for labeled images and reduce labor costs. In addition, due to the use of unlabeled images for training and fine-tuning the model, our model is more adaptable to the images obtained from different hospitals than trained by supervised learning. Therefore, our method can be more widely applied to different datasets.

## 2. Related work

Current studies only used supervised learning methods for PE lesion segmentation. Carlos et. al. used 2D, 2.5D, and 3D U-Net models to train PE CTPA images and compared the differences between these models [5]. The sensitivity of 3D U-Net, which performed the best, was 0.55. Kun et. al. used mask RCNN for object detection on PE CTPA images [4] and achieved a sensitivity of 0.747. However, the above-mentioned studies were all tested on a single dataset and the model only performed well on the dataset used in above-mentioned studies. The CTPA images vary according to the machine settings of different hospitals. Therefore, this study proposed to use semi-supervised learning methods to improve the model adaptability of different datasets, so that the model can be immediately applied to CTPA images obtained from different hospitals, while the accuracy is maintained.

Semi-supervised learning methods were also used in the magnetic resonance imaging (MRI) and microscope images. Dong et. al. used an adversarial semi-supervised learning architecture to train MRI images for segmenting prostate, bladder, and rectum [8]. By adding unlabeled images during training, the model can revise the predicted boundary. Zhou et. al. applied semi-supervised learning to microscopic images for annotating diabetic retinopathy [9]. It was verified that semi-supervised learning can also perform well in small objects segmentation with complex shapes. Therefore, we proposed to apply semi-supervised learning to the PE dataset, which contains small objects, and enable our model to better adapt to different datasets and improve the accuracy.

## 3. Methods

*3.1 Feature-enhanced adversarial semi-supervised segmentation*

Our semi-supervised semantic segmentation architecture referred to the adversarial learning architecture by Hung et. al [10]. The concept is to train a segmentation network and a discriminator with labeled data, so that the discriminator can be used to provide prediction with positional features when predicting unlabeled data. The model Hung et al. used is mainly for panoptic segmentation. The input of the discriminator was the masks of different classes, which enabled the discriminator to learn features for differentiating different classes. However, there is only one class in the PE task, making the discriminator hard to learn features through a single mask. To better leverage the discriminator in the PE task, we proposed a novel design by adding the feature map, which was extracted from the semantic segmentation model encoder, to the input of the discriminator, as shown in Fig 2. Therefore, the discriminator no longer learned from the masks, but from the mask position and PE-related features. Our semi-supervised segmentation network contained a segmentation network and a discriminator. The segmentation network is a revised version of HRNet with enhanced PE features in order to increase computation efficiency while maintaining high-resolution convolutional operations. The added discriminator trained by adversarial learning methods can further make the prediction of unlabeled data close to the real data annotation.

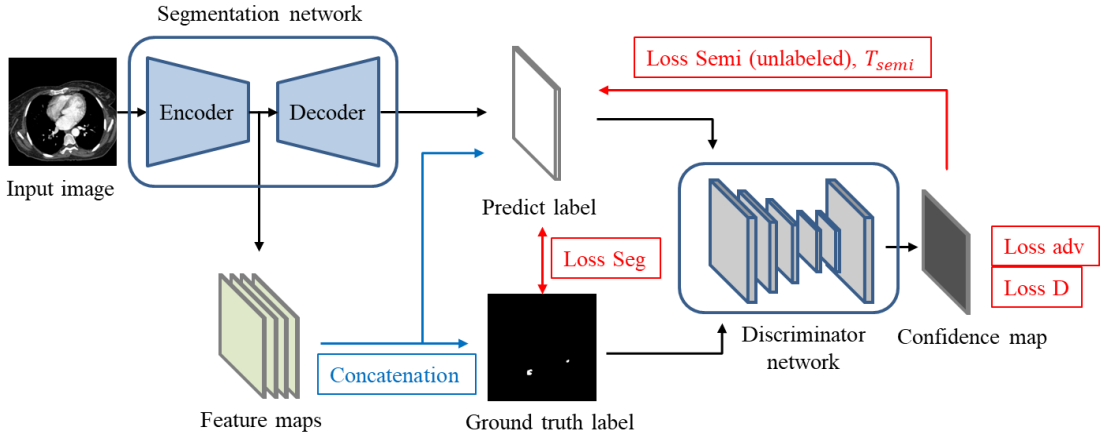

Fig 2. Feature-enhanced adversarial semi-supervised segmentation architecture. The segmentation network and the discriminator network were trained by labeled data first. Then, the unlabeled data are used to fine-tune the segmentation network. The feature maps generated from the encoder were used as the input of discriminator to improve the ability of discriminator.

*3.1.1 Segmentation network*

Although HRNet can extract features with high resolution, it is still not enough for segmenting the small-sized PE features. Therefore, we modified our HRNet-based segmentation model to extract the features of PE at higher resolution. To increase computation efficiency while detecting small objects at high resolution, we made three modifications to HRNet save computing resources while detecting small PE objects at a higher resolution, as shown in Fig 3. The purpose of the modification was to maintain the PE feature at a high resolution while training. Moreover, we used the CARAFE upsampling module to replace the bilinear upsampling module of the encoder to make the predicted shapes and boundaries more accurate. The first modification was to remove the second convolutional layer of HRNet, so that the downsampling rate before entering the network was reduced from 1/4 to only 1/2. This allows the images to preserve a larger feature map, which is suitable for PE feature detection. However, it would cause a significant increase of the computing cost. Therefore, in the second modification, we removed the 4$^{th}$ stage of the original HRNet to reduce the computation complexity. Since the feature map of the smallest scale size in 4$^{th}$ stage is only $25 \times 25$ (pixels), the boundary and location features of PE are found to be blurred via the visualized feature map, which made the decoder hard to improve the prediction accuracy effectively. Therefore, once the 4$^{th}$ stage was removed, the computing cost can be greatly reduced while the accuracy remained. The third modification is to build the CARAFE module as the upsampling layer of the decoder. Compared with bilinear interpolation, the CARAFE module predicted the upsampling kernel for each pixel in the feature map, so that it can better restore the shape and position of the PE.

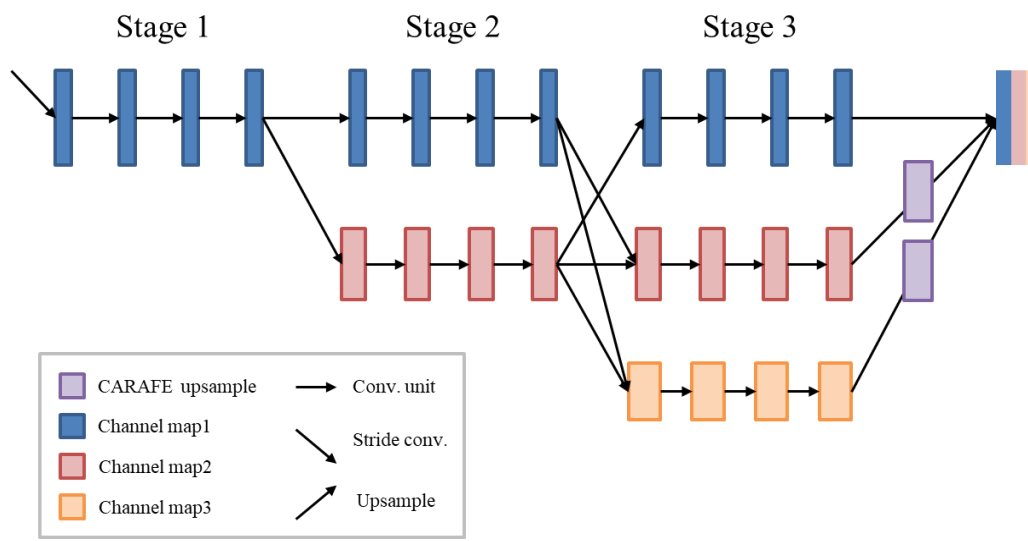

Fig 3. Segmentation network structure. We used HRNet as our backbone. Unlike the original structure of HRNet, we only downsampled the image 3 times to make the feature map clearer. The CARAFE module was used for upsampling in the decoder to help restore the predicted position and shape.

*3.1.2 Discriminator network*

The discriminator network is the core of adversarial semi-supervised learning. The discriminator network was used to generate the confidence maps from predicted mask and ground truth mask for adversarial learning. Two loss functions were used to train the discriminator. Loss_D was used to differentiate the ground truth confidence map from the predicted confidence map. Loss_adv was used to calculate the similarity of the ground truth confidence map and the predicted confidence map. By minimizing the sum of loss functions, the discriminator can make the predicted mask closer to ground truth map. Thus, the discriminator can revise the predicted mask while training with unlabeled PE images. As shown in Fig 4, the discriminator consisted of three convolutional layers and an upsampling layer. As the input of the discriminator, the feature map generated by the encoder of segmentation network was concatenated with the prediction mask and the weight of the prediction mask was set to be 5. Therefore, the discriminator focused on both the positional information in the mask and PE features. Through our proposed discriminator, the mask of unlabeled PE images can achieve higher accuracy.

*3.2 Post-processing*

In order to ensure that the prediction results were inside the lung, we added two correction processes after the output of our semi-supervised segmentation model: lung area mask and boundary lines. For the lung mask, the CTPA image HU threshold was set to -160 to binarize the image. Then, in the middle of the cropped area, a pixel with 0 value was selected as the starting point for region growing. Finally, the lung mask grown by region growing was used as a filter for predicted mask, so that we can focus PE lesions on the lung area. However, since the CTPA images obtained from different hospital differed, the PE images had different size, brightness, and displacement, the boundary lines method was also used. We first binarized the image by setting the HU threshold to -160 and found the top and bottom boundary of the lungs. Then, the area which was out of the boundary lines were removed from the predicted mask. The boundary lines method was used after the lung area mask methods. By adding the two correction processes, we can ensure most of our predicted masks were inside the lung area.

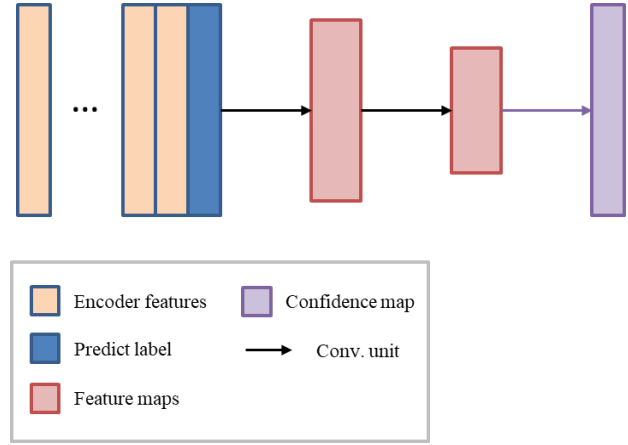

Fig 4. Discriminator network structure. Five convolutional layers are used. Output was a confidence map, which can help determine whether the predicted mask of unlabeled PE image was correct.

**4. Experiments and results**

Our semi-supervised segmentation model was trained with a labeled dataset and an unlabeled dataset. The prediction accuracy of the unlabeled dataset was increased after the training. Our model was also applied to other PE datasets. The results showed that only a small amount of unlabeled data was required for fine-tuning and the prediction accuracy of unlabeled data could be greatly improved by our method. In this study, we used two different datasets to train the model and one external dataset to evaluate the model. The mIOU, dice score, and sensitivity were used for evaluation. Furthermore, our model was compared with the 2D and 2.5D U-Net models proposed by Carlos et. al using supervised learning only.

*4.1 Training and evaluation datasets*

In this study, three different CTPA datasets were used, including the open source dataset, the National Cheng Kung University Hospital (NCKUH) dataset, and the China Medical University Hospital (CMUH) dataset. (I) The open source dataset was provided by Mojtaba et. al [11]. A total of 33 patients were included in the dataset, 27 patients in training set and 6 patients in validation set, and all the data were labeled images. This dataset was used to train the supervised segmentation model as pre-training. (II) The second dataset was provided by the NCKUH, with 70 patients. The NCKUH dataset included 63 patients in training set (unlabeled images) and 7 patients in validation set (labeled images). Both the training sets of NCKUH dataset and open source dataset were jointly used to train our semi-supervised segmentation model and the two validation sets were used to evaluate the trained model. (III) The third dataset was provided by the CMUH, which included 27 patients and were used as the external testing sets in this study. We used a small amount of CMUH unlabeled images (5 patients) to fine-tune our model and then used the remaining images of 22 patients (labeled images) for testing sets. The results were compared with the ground truth masks manually annotated by the physician. Both NCKUH dataset and CMUH dataset were composed of axial images from chest CT scans performed using a pulmonary angiography protocol. The annotation of the two datasets were reviewed by board-certificated radiologists and contoured by board-certificated radiation oncologist. This study was approved by institutional review board of NCKUH and CMUH.

*4.2 Implementation details*

In this study, the training process of semi-supervised segmentation model can be divided into two steps. First, we trained the supervised segmentation model. Then, we used the parameters of the trained model as the pre-trained parameters for the semi-supervised model. When training the supervised segmentation model, we used only the open source dataset, which was labeled images. A total of 100 epochs were trained. The loss function was binary cross entropy (BCE) loss and dice loss. The optimizer was stochastic gradient descent (SGD) [12] with momentum 0.9. The initial learning rate was set to 1e-4. For semi-supervised segmentation, both the open source dataset and NCKUH dataset were used during training. In the segmentation network, BCE loss and dice loss with SGD optimizer were also used. The initial learning rate was set to 2e-4. In the discriminator network, BCE loss was used and the initial learning rate was set to 1e-4 due to its simple architecture.

*4.3 Open source dataset supervised model results*

The pre-trained model for semi-supervised learning were trained by open source dataset. In order to demonstrate the ability of our segmentation model, we compared our proposed segmentation model with other baseline models using only supervised learning. In this study, we tested the basic segmentation models, including 2D U-Net [13], 2.5D U-Net, DeepLabV3+ [14], and HRNet, on the open source dataset. The results are shown in Table 1. The 2D U-Net performed better than the 2.5D U-Net on our dataset, which was consistent with the results of Carlos et. al. The HRNet performed better on PE images because it retained high resolution architecture for convolutional operations and integrated feature maps of various scales. We further improved our HRNet-based model by increasing the input size (HRNet large), adding the CARAFE module (HRNet Large/CARAFE), and removing the 4$^{th}$ stage in HRNet (HRNet Large/CARAFE/without stage4). The resulting mIOU, dice score, and sensitivity achieved 0.4801, 0.6018, and 0.6801, respectively. Since the 4$^{th}$ stage was removed from HRNet, our model's performance was slightly lower than that of the HRNet with 4[th] stage in supervised learning. However, the HRNet without 4[th] stage obtained higher computing efficiency during semi-supervised learning in section 3.4. The visualization the predicted results was shown in Fig 5. When the lesion area of PE was small, the prediction was usually larger than ground truth. Hence, the false positives lower the accuracy significantly. Nevertheless, the false positives were not so concerned in clinical diagnosis. Thus, we also evaluated our model by dice score and sensitivity, which focus on true positive area. The high true positives showed that our model can correctly annotate the PE lesion at most of the time.

Table 1. Supervised segmentation results of open source dataset. HRNet (Large/CARAFE) achieved the best mIOU, but HRNet (Large/CARAFE/without stage4) had less parameter which is more suitable for semi-supervised learning.

| Supervised segmentation (Open source dataset) | | | |
|---|---|---|---|
| Segmentation model | Validation set mIOU | Dice score | Sensitivity |
| 2D U-Net | 0.3433 | 0.4168 | 0.4831 |
| 2.5D U-Net | 0.3378 | 0.4061 | 0.4712 |
| DeepLabv3+ | 0.3044 | 0.3707 | 0.4455 |
| HRNet | 0.4218 | 0.5168 | 0.5750 |
| Our (HRNet Large) | 0.4611 | 0.5877 | 0.6735 |
| Our (HRNet Large/CARAFE) | **0.4801** | **0.6018** | 0.6810 |
| Our (HRNet Large/CARAFE/without stage4) | 0.4683 | 0.5938 | **0.6856** |

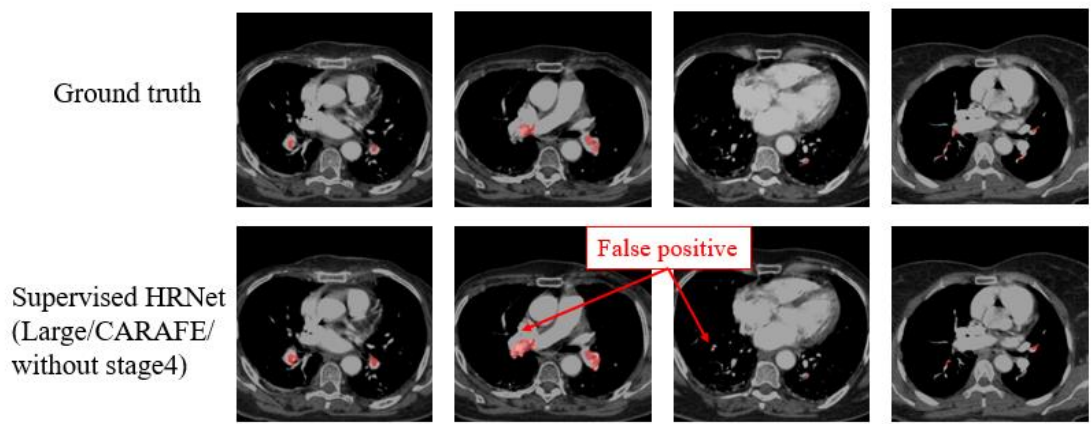

Fig 5. Predicted images of supervised segmentation in the open source dataset. The first column is the ground truth and the second column is our predictions. The false positive area was shown in red color.

*4.4 Semi-supervised results*

We tested the supervised learning model on the NCKUH dataset and CMUH dataset. The results showed that the supervised learning model overly relied on the PE features of the training dataset, so performance on these two datasets were not ideal. In order to adapt our model to different datasets, we added the unlabeled PE images from NCKUH for semi-supervised learning. Besides, we also used an additional CMUH dataset to fine-tune and test our semi-supervised model. In this section, we used three datasets to evaluate our semi-supervised segmentation model and compared the differences between supervised learning and semi-supervised learning in predicting unlabeled images. Through semi-supervised learning, our model can achieve great improvements in accuracy on different datasets by fine-tuning with a small number of unlabeled images.

*4.4.1 Open source dataset semi-supervised model results*

In the present study, we used HRNet (Large/CARAFE/without stage4) as the backbone of the semi-supervised segmentation network. We also tested the semi-supervised models with different segmentation backbones. As shown in Table 2, the HRNet-based models can achieve higher accuracy than U-Net and DeepLabV3+. Since the size of PE lesion was small, a higher-resolution convolutional operation was needed to extract PE features. Comparing to the original HRNet, our proposed model enhanced the feature extraction capability for small objects, so that the accuracy was greatly improved. In order to increase the PE mask identification ability of discriminator, we concatenated feature maps with the prediction and ground truth mask as the input of the discriminator. Although the mIOU was not improved significantly, the dice score and the sensitivity were improved. As shown in Fig 6, after adding feature maps, the model can better revise the predicted shape and boundary. This results showed that our feature-enhanced methods can increase true positives and improve dice score and sensitivity.

Table 2. Semi-supervised results of open source dataset. The mIOU, dice score, and sensitivity were improved after adding feature maps before the discriminator.

| Semi-supervised segmentation (Open source dataset) | | | |
|---|---|---|---|
| Segmentation backbone | Validation set mIOU | Dice score | Sensitivity |
| U-Net | 0.2810 | 0.3418 | 0.4927 |
| DeepLabv3+ | 0.2513 | 0.3188 | 0.4205 |
| HRNet | 0.3142 | 0.4323 | 0.5764 |
| Our (HRNet Large/CARAFE/ without stage4) | 0.4548 | 0.5588 | 0.7056 |
| Our (HRNet Large/CARAFE/ without stage4) + feature maps | **0.4549** | **0.5611** | **0.7180** |

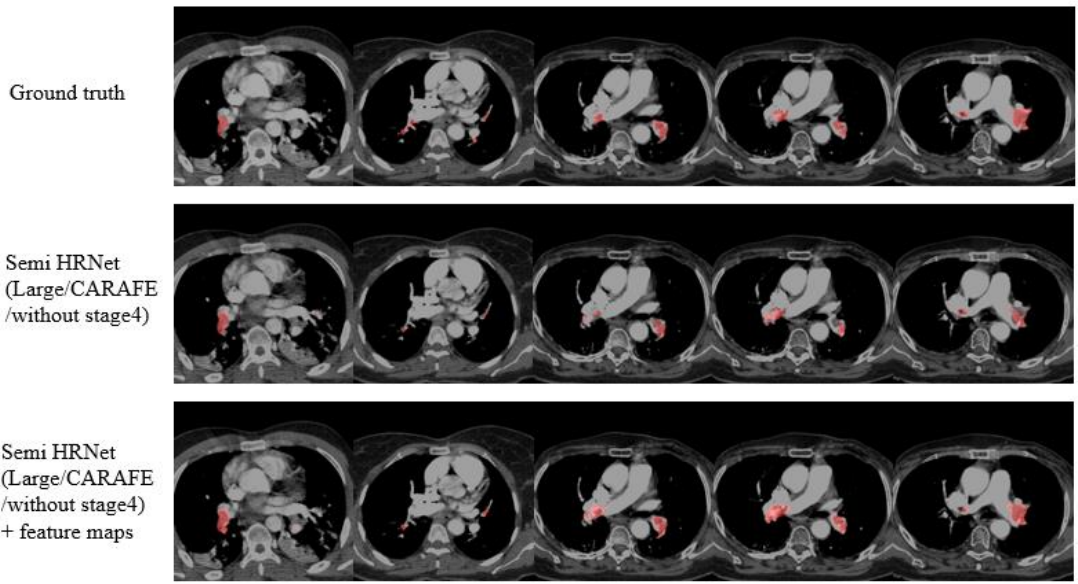

Fig 6. Predicted images of semi-supervised segmentation in open source dataset. The second column was our model without adding feature maps before the discriminator. The third column showed that adding the feature maps before the discriminator can reduce false negatives.

The comparison of semi-supervised learning results and supervised learning results for the open source dataset was shown in Table 3. The mIOU and the dice score of the semi-supervised learning method slightly decreased, but the sensitivity has been significantly improved. The result showed that semi-supervised learning can increase true positives. However, the semi-supervised model tended to predict the PE lesion area larger, which in turn leaded to an increase in false positives. Therefore, the sensitivity was much higher than mIOU and dice score. By comparing the results of supervised learning and semi-supervised learning on the open dataset, we found that semi-supervised learning had limited improvement on the training dataset, but it can revise the predicted boundary to improve the true positives.

Table 3. Comparison between supervised and semi-supervised of open source dataset. The higher sensitivity showed that semi-supervised method had higher true positives.

| | Segmentation backbone | Open source dataset | | |
|---|---|---|---|---|
| | | Validation set mIOU | Dice score | Sensitivity |
| Supervised segmentation | Our (HRNet Large/CARAFE/without stage4) | **0.4683** | **0.5938** | 0.6856 |
| Semi-supervised segmentation | Our (HRNet Large/CARAFE/without stage4) | 0.4548 | 0.5588 | 0.7056 |
| | Our (HRNet Large/CARAFE/without stage4) + feature maps | 0.4549 | 0.5611 | **0.7180** |

*4.4.2 NCKUH dataset semi-supervised model results*

We used the NCKUH dataset as the unlabeled training dataset. In order to evaluate our semi-supervised model on unlabeled images, we segmented a validation set from the NCKUH dataset to evaluate both the supervised model and semi-supervised model. Besides, the NCKUH dataset was not used for fine-tuning while testing the supervised model. As shown in Table 4, if the supervised model was not fine-tuned for other datasets, the results were not ideal. However, through semi-supervised training, the accuracy of unlabeled image was significantly improved with only a small amount of additional unlabeled images. Using our feature-enhanced adversarial semi-supervised segmentation model, the resulting mIOU, dice score, and sensitivity could achieve 0.3510, 0.4854 and 0.4253, respectively. The results showed that semi-supervised learning can improve the accuracy of unlabeled images by adding unlabeled dataset to training. As shown in Fig 7, the vascular imaging of NCKUH dataset was more obvious than imaging of open source dataset. Therefore, the supervised model trained on the open source dataset was not able to perform well on the NCKUH dataset, while the semi-supervised learning can solve the problem of over-fitting to specific dataset in supervised learning.

Table 4. Comparison between supervised and semi-supervised model on NCKUH dataset. Through semi-supervised learning, the accuracy of unlabeled data can be significantly improved.

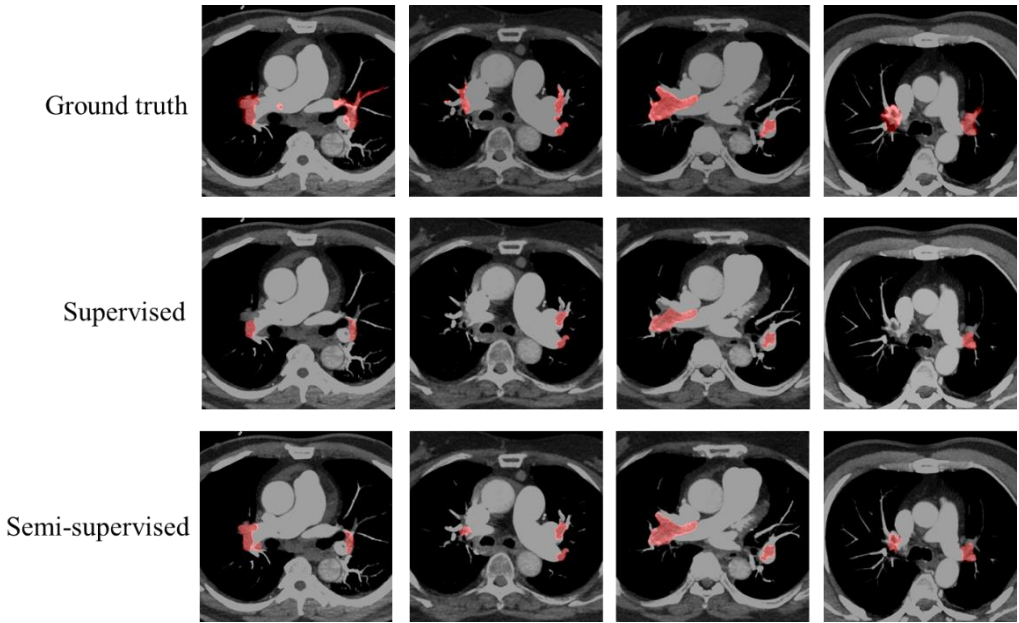

| | Segmentation backbone | Testing set mIOU | Dice score | Sensitivity |
|---|---|---|---|---|
| Supervised segmentation | HRNet (Large/CARAFE/without stage4) | 0.2128 | 0.3488 | 0.3169 |
| Semi-supervised segmentation | HRNet (Large/CARAFE/without stage4) | 0.3386 | 0.4597 | 0.4088 |
| | HRNet (Large/CARAFE/without stage4) + feature maps | **0.3510** | **0.4854** | **0.4253** |

NCKUH dataset

Fig 7. Predicted images of segmentation result in NCKUH dataset. Same as the results of open source dataset, the false negatives are reduced after semi-supervised training.

*4.4.3 CMUH dataset semi-supervised model results*

In order to verify whether our semi-supervised model can be applied to different datasets, we used the CMUH dataset for additional testing. We only added a small amount of CMUH unlabeled images (5 patients) to fine-tune the model. From Table 5, through fine-tuning of semi-supervised learning, the resulting mIOU, dice score, and sensitivity improved from 0.2344, 0.3325, and 0.3151 to 0.3721, 0.5113, and 0.4967, respectively. As shown in Fig 8, the semi-supervised model can fill the PE lesion area more completely than supervised model. This result showed that our semi-supervised model can work well on other unlabeled datasets.

Table 5. Comparison between supervised and semi-supervised of CMUH dataset. The accuracy was improved through fine-tuning the semi-supervised model with a small amount of CMUH images.

| CMUH dataset | | | | |
|---|---|---|---|---|
| Segmentation backbone | | Testing set mIOU | Dice score | Sensitivity |
| Supervised segmentation | HRNet (Large/CARAFE/without_stage4) | 0.2344 | 0.3325 | 0.3151 |
| Semi-supervised segmentation | HRNet (Large/CARAFE/without_stage4) + feature maps | **0.3721** | **0.5113** | **0.4967** |

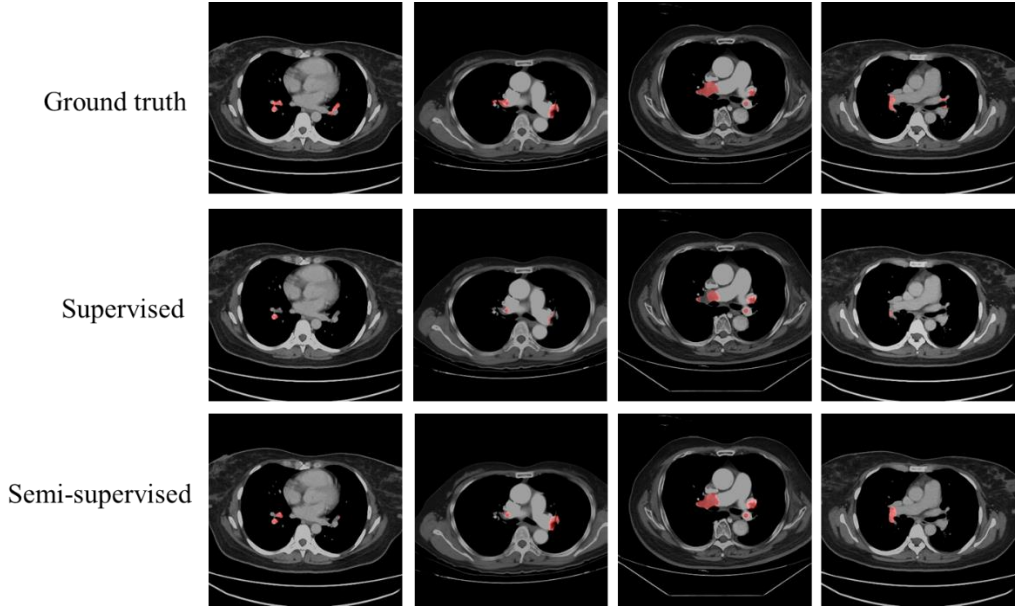

Fig 8. Predicted images of segmentation result in CMUH dataset. By fine-tuning the semi-supervised model with images of 5 patients, our model can better fit the shapes in the PE lesion area.

## 5. Discussion

In this work, we first designed a supervised segmentation network which is suitable for the PE dataset. Different from the models of Carlos et. al, we used a HRNet-based model that is more suitable for small object detection. Therefore, our supervised model can achieve better dice score and sensitivity. Then, in order to better fit our model to different PE datasets, we designed a feature-enhanced adversarial semi-supervised segmentation model. We can find that the semi-supervised model did not significantly improve the accuracy on the testing set of open source dataset. The improvement was limited due to the used of the same training set as supervised learning. Therefore, our semi-supervised model can just slightly revise the predicted boundary and shape. However, our semi-supervised learning can achieve a significant improvement in the accuracy of the NCKUH dataset and the CMUH dataset. By adding the feature maps to discriminator for training, our model can mark the shape of the PE lesion area more accurately and thus increased the accuracy.

We can find that the sensitivity was higher than the dice score only on the open source dataset. The reason is that the prediction of the open source dataset contained more false positives, which is unfavorable for the calculation of dice score. However, the predictions of other datasets rarely contained false positives, so the dice score was slightly higher than the sensitivity. This difference was mainly due to the reason that the open source dataset and other datasets had different labeling personnel, resulting in different labeling standards. The open source dataset had stricter standards for the PE lesion area and only the obstructed area was annotated. In other datasets, all areas around the obstructed area were annotated. Therefore, the true positives were increased and the false positives were decreased in the NCKUH dataset and the CMUH dataset.

The prediction results of unlabeled images (NCKUH and CMUH datasets) were shown in Fig 9. The complex shapes and the large number of PE lesion area resulted in the decrease in accuracy. Our model can only predict the approximate location of these images and cannot fully present the details, resulting in a large number of false negative pixels. However, the approximate location we predicted was helpful enough for doctors to confirm the PE lesion area. Therefore, despite the existence of a large number of false negative pixels, our model can still assist doctors in quickly finding the PE lesion area.

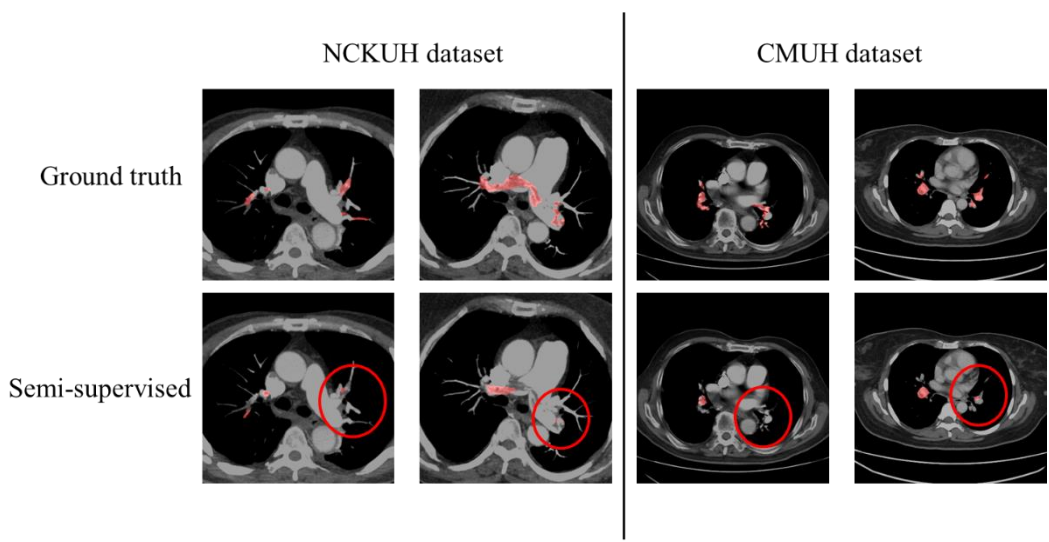

Fig 9. Comparison of NCKUH and CMUH dataset. There were lots of false negative pixels in our semi-supervised prediction. The false negative pixels make the accuracy decrease.

## 6. Conclusion

In this work, we proposed a semi-supervised semantic segmentation model for PE lesion annotation. In the semantic segmentation network, we increased the size of the input feature map to preserve the features of PE. We also made appropriate modifications to the oversized model to reduce computational consumption while maintaining high accuracy. The novelty of our model is to add the feature information to the discriminator by using the feature map of the encoder to improve the accuracy.

In the open source dataset, our supervised model surpassed the model of Carlos et. al. The resulting mIOU, dice score, and sensitivity achieved 0.4683, 0.5938, and 0.6856, respectively. Besides, by adding a small amount of unlabeled PE images for fine-tuning the model, the resulting mIOU, dice score, and sensitivity of the NCKUH dataset could be improved to 0.3510, 0.4854, and 0.4253. The resulting mIOU, dice score, and sensitivity of the CMUH dataset could also be improved to 0.3721, 0.5113, and 0.4967, respectively. Superior to the result of supervised learning, our semi-supervised model is more accurate in predicting unlabeled images. It also showed that our semi-supervised model can be applied to other PE datasets.

For CTPA images obtained from different hospitals, we can fine-tune our model by adding a small amount of unlabeled images to achieve an accuracy equivalent to the accuracy of the training dataset. Although the accuracy of our semi-supervised learning model may still be improved, we have demonstrated a promising training method that can be easily applied to multiple datasets. In the future, our model may be applied to other medical imaging datasets to reduce the labor cost of labeling and facilitate medical imaging research.